\documentclass[letterpaper,english,aps, nofootinbib, pr, twocolumn, lengthcheck, superscriptaddress]{revtex4-2}

\usepackage[T1]{fontenc}
\usepackage[colorlinks=true,linktocpage=true,linkcolor=blue,citecolor=blue]{hyperref}
\usepackage{color,txfonts}
\usepackage{verbatim}
\usepackage{bm}
\usepackage{amssymb}
\usepackage{graphicx}
\usepackage{slashed}
\usepackage{wrapfig}
\usepackage{bbm}
\usepackage[caption=false]{subfig}
\usepackage{verbatim}

\begin{document}

\title{Merger and post-merger of binary neutron stars with a quark-hadron crossover equation of state}

\author{Yong-Jia Huang$^{1,2,3}$, Luca Baiotti$^{4}$, Toru Kojo$^{5,6}$, Kentaro Takami$^{7,3}$, Hajime Sotani$^{8,3}$, Hajime Togashi$^6$, Tetsuo Hatsuda$^3$, Shigehiro Nagataki$^{3,8}$ and Yi-Zhong Fan$^{1,2}$}
\noaffiliation

\affiliation{
Key Laboratory of Dark Matter and Space Astronomy, Purple Mountain Observatory, Chinese Academy of Science, Nanjing, 210023, China.}
\affiliation{
School of Astronomy and Space Sciences, University of Science and Technology of China, Hefei, Anhui 230026, China.}
\affiliation{RIKEN Interdisciplinary Theoretical and Mathematical Sciences Program (iTHEMS), RIKEN, Wako 351-0198, Japan.}
\affiliation{International College and Graduate School of Science, Osaka
  University, 1-1 Machikaneyama, Toyonaka 560-0043, Japan}
\affiliation{Key Laboratory of Quark and Lepton Physics (MOE) and Institute of Particle Physics, Central China Normal University, Wuhan 430079, China}
\affiliation{Department of Physics, Tohoku University, Sendai 980-8578, Japan}
\affiliation{Kobe City College of Technology, 651-2194 Kobe, Japan}
\affiliation{RIKEN Astrophysical Big Bang Laboratory (ABBL), Cluster for Pioneering Research, Wako, Saitama 351-0198, Japan}

\begin{abstract}

Fully general-relativistic binary-neutron-star (BNS) merger simulations with quark-hadron crossover (QHC) equations of state (EOSs) are studied for the first time.
In contrast to EOSs with purely hadronic matter  or with a  first-order
quark-hadron phase transition (1PT), in the transition region QHC EOSs show a peak in sound speed, and thus a stiffening. We study the effects of such stiffening in the  merger and post-merger gravitational (GW) signals.  Through simulations in the binary-mass range $2.5 < M/M_{\odot} < 2.75$, characteristic differences due to different EOSs appear in the frequency of the main peak of the post-merger GW spectrum ($f_2$), extracted through Bayesian inference. In particular, we found that
  (i) for lower-mass binaries, since the maximum baryon number density ($n_{\rm max}$) after the merger stays below $3\text{--}4$ times the nuclear-matter density ($n_0$), the characteristic stiffening of the QHC models in that density range results in a lower $f_2$ than that computed for the underlying hadronic EOS and thus also than that for EOSs with a 1PT, (ii) for higher-mass binaries, where $n_{\rm max}$ may exceed $4\text{--}5 n_0$ depending on the EOS model, whether $f_2$ in QHC models is higher or lower than that in the underlying hadronic model  depends on the height of the sound-speed peak.
Comparing the values of $f_2$ for different EOSs and BNS masses gives important clues on how to discriminate different types of quark dynamics in the high-density end of EOSs and is relevant to future kilohertz GW observations with third-generation GW detectors.

\end{abstract}


\maketitle

\begin{figure}[htbp]
\centering
\vspace{-0.3cm}
\includegraphics[width=0.45\textwidth]{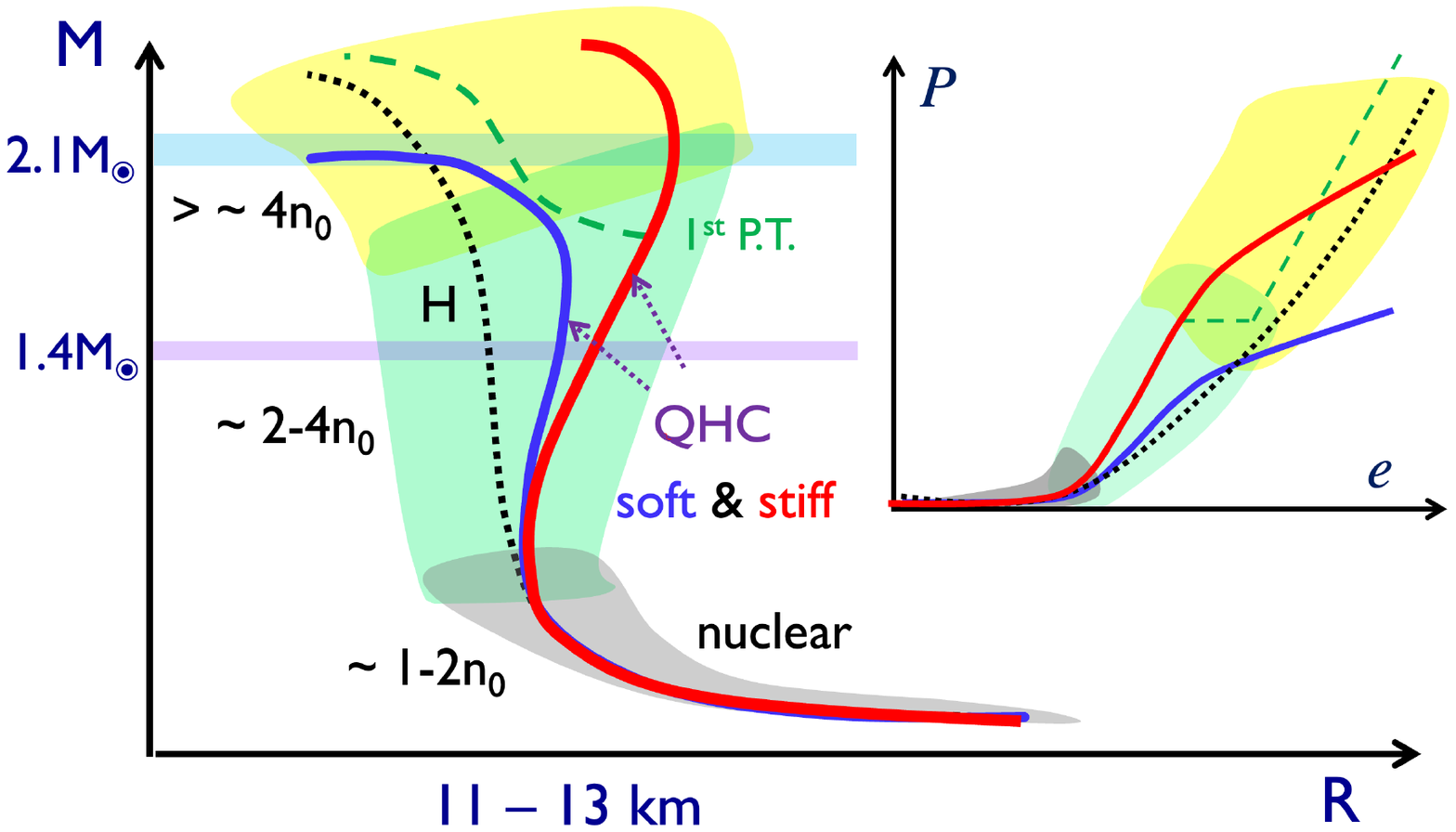}
\vspace{-0.3cm}
\caption{Schematic plots for the mass-radius relations (main figure) and pressure $P$ vs. energy density $e$ (inset) for some EOSs satisfying constraints from terrestrial experiments and the observational fact that a NS of mass $\approx 2M_\odot$ exists; ``H" refers to a purely hadronic model, "1PT" to a hybrid model with a first-order quark-hadron phase transition, and ``QHC" to models with a quark-hadron crossover. QHC models show stiffening at densities lower than in the other two cases, typically leading to larger radii and smaller central densities for NSs with masses $1.4\text{--}2M_\odot$.
 The grey, green, and yellow shaded areas in both the main figure and inset correspond to the number density ranges of $n \sim 1 \text{--}2n_0$, $\sim 2\text{--}4n_0$, and $\gtrsim$ $4n_0$, respectively.
 }
\label{fig:schematic}
\end{figure}

{\it Introduction.} Multi-messenger astronomy, including gravitational,
electromagnetic, and neutrino signals, has started offering new ways of
obtaining information  on ultra-high-density matter
\cite{LIGOScientific:2017vwq}. Observations of the inspiral of a binary
neutron-star (BNS) merger may provide information on the equation of
state (EOS) at a few times the nuclear saturation number density ($n_0 =
0.16\,{\rm fm}^{-3}$), and even higher densities (several times $n_0$)
may be investigated through observations of the post-merger phase, where
matter is also hotter than in the inspiraling NSs \cite{baiotti2019,
  Burgio2021}. In the near future, the detection of BNS mergers will
happen on a daily basis and this will also allow one to perform improved statistical analyses of the properties of their EOS.

The EOS plays a crucial role in determining the structure of NSs. See Fig.\ref{fig:schematic} for schematic plots of the mass-radius and energy-pressure relations in NSs with different types of EOSs.
Nuclear EOSs based on microscopic nuclear two- and three-body forces are supposed to be valid up to number densities $n \simeq 1.5\text{--}2n_0$, and thus to describe somewhat accurately the equatorial radii of canonical NSs (mass $M \simeq 1.4M_\odot$), which have core densities around $2\text{--}3n_0$. The most massive NS known, PSR J0740+6620, has mass $M/M_\odot = 2.08\pm 0.07$ \cite{Fonseca:2021wxt} and inferred core density $\gtrsim 3\text{--}4n_0$. This is close to the density at which  baryons with radii $\simeq 0.5\text{--}0.8$ fm begin to overlap, presumably resulting in matter beyond the purely hadronic regime, such as  quark matter.

One of the fundamental questions in the study of ultra-dense matter is
how the quark-hadron phase transition takes place. The most intensively
studied scenario is the one involving first-order quark-hadron phase transitions (1PTs). In this case, it is usually believed that pressure support (and thus the radius of the material object resulting from the merger) decreases abruptly after the phase transition. 
Such a change in compactness would appear, in turn, as a (possibly measurable) shift to higher values of the frequency of gravitational waves (GWs) emitted from the merged object \cite{Bauswein2019, Most:2018eaw, Most2020, weih2020, Blacker2020, Liebling2021, Prakash2021}.
A too large reduction of the stellar radius, however, is disfavored by the recent NICER observations and analyses, reporting similar radii for NSs with masses of $1.4M_\odot$ and $2.1M_\odot$ \cite{Miller:2021qha, Riley:2021pdl, Raaijmakers:2021uju, Han2022}. 

\begin{figure}[hbtp]
\centering
\includegraphics[width=0.43\textwidth]{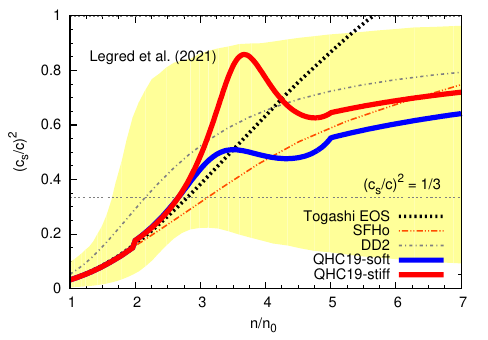}
\caption{Square of sound speed normalized to the speed of light, $c_s^2/c^2 = {\rm d}  P/{\rm d} e$, for our QHC EOSs with soft and stiff sets of quark model parameters and for representative hadronic EOSs: Togashi EOS \cite{togashi2017}, SFHo \cite{Steiner:2012rk}, and DD2 \cite{DD2}.
The yellow band is the allowed region in the model-agnostic approach of
Legred {\it et al.} \cite{Legred:2021hdx} (see also \cite{Han2021}).
The conformal limit, $c_s^2=c^2/3$, which should be reached in the high-density limit, is also shown as a guide.
}
\label{fig:cs2-nb}
\end{figure}

An alternative to a 1PT is a continuous crossover from hadronic matter to quark matter. Some of the present authors constructed quark-hadron-crossover (QHC) EOSs \cite{Masuda:2012kf,Baym:2017whm,baym2019,kojo2021b}, generally finding a peak in the sound speed, $c_s/c = \sqrt{ {\rm d} P/{\rm d} e}$, exceeding the conformal limit $c/\sqrt{3}$, with $c$ being the speed of light; see Fig.\ref{fig:cs2-nb}. Microscopic considerations on the structure of such a peak \cite{McLerran:2018hbz,kojo2021a} emphasize the importance of quark substructure in baryons and of quark Pauli blocking effects. 
Peaks in sound speed are absent in EOSs involving 1PTs or in purely hadronic models. In the latter,  stiffening results from nuclear many-body repulsions, which keep growing with density, leading to monotonic growth in sound speed. The existence of a  peak in sound speed in QHC EOSs is unique and can be taken as the signature for the onset of quark-matter formation.

\begin{figure}[hbtp]
    \centering
	\includegraphics[width=0.52 \textwidth]{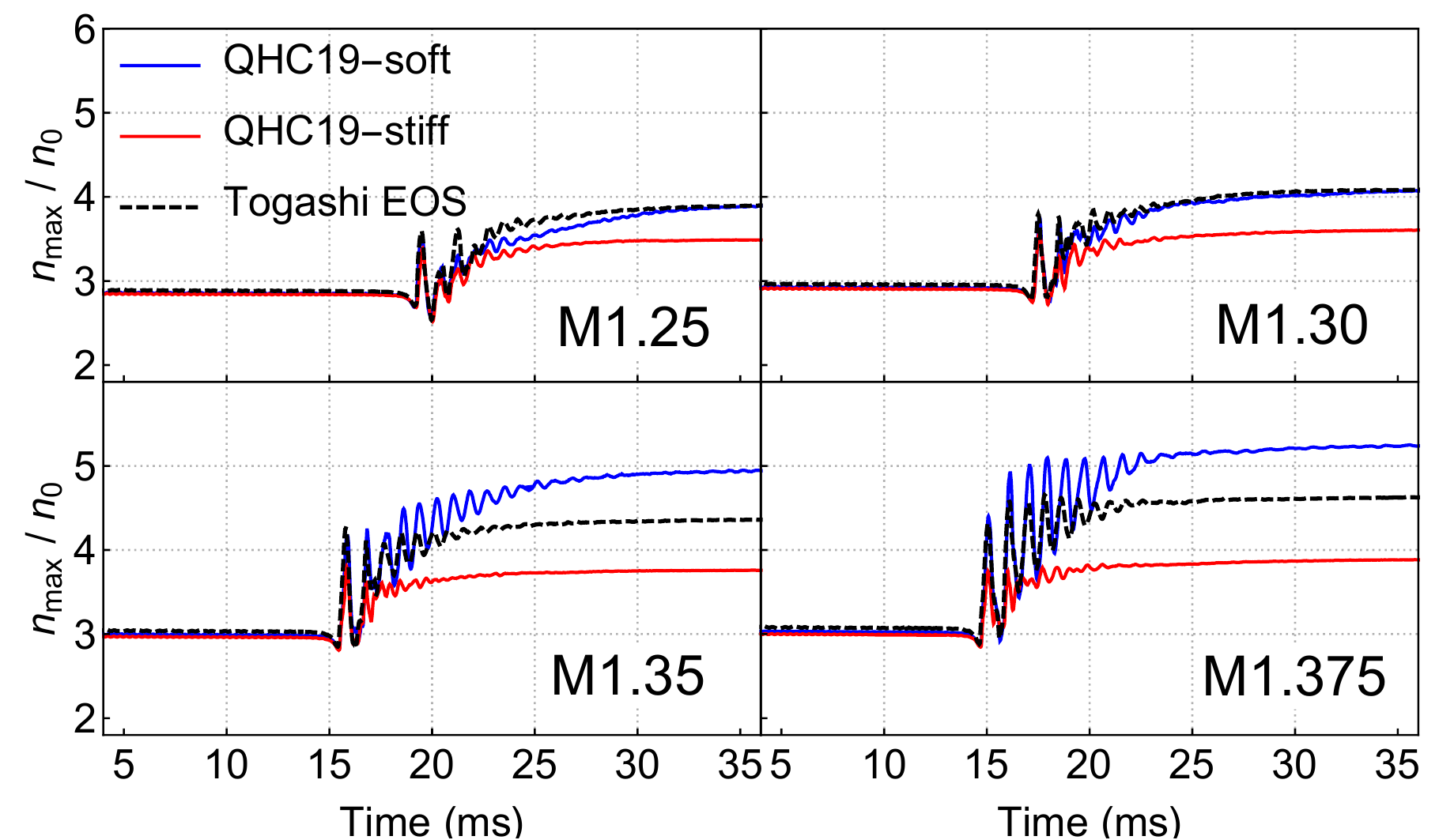}
    \caption{Evolution of the maximum number density for simulations employing the QHC19 and Togashi EOSs with different initial masses.
    $t_{\rm merger}$ is the merger time defined as the time of the maximum amplitude of $|h| \equiv (h_{+}^2 + h_{\times}^2)^{1/2}$.
    } 
    \label{fig:n_max}
\end{figure}

In this Letter, for the first time, results of numerical simulations of
BNS mergers with  EOSs based on QHC are reported. We adopt the QHC19 EOS
\cite{baym2019}, which is based on the Togashi nucleonic EOS
\cite{togashi2017} for $n \le 2n_0$ and a pure quark EOS for $n\gtrsim
5n_0$, with the crossover region calculated through interpolation
\cite{baym2019}. We compare results with simulations adopting the Togashi
EOSs over the whole density range. The QHC19 and Togashi EOSs differ
substantially only for $n \gtrsim 3n_0$, and, since the maximum values of
$n$ in our inspiraling NSs are around $3n_{0}$ ({\it cf.}
Fig.~\ref{fig:n_max}), the properties (like tidal deformability
\cite{hinderer2007,hinderer2009}) of stars built with the above different
EOSs and their dynamics (like the evolution of the central number density
or of the GW frequency) during the inspiral differ by less than 1\% (see
Table 1 in the Supplemental Material). 
More remarkable differences are expected only during and after the merger.

{\it Numerical setup.} 
As a first step to explore the role of a QHC in BNS mergers, we focus on equal-mass configurations, and, with the goal of studying post-merger dynamics, we chose four relatively low-mass models, in which the gravitational masses of each NS at infinite separation are $M/M_{\odot} = 1.250,~1.300,~1.350,~1.375$. We refer to these as M1.25, M1.30, M1.35, M1.375, respectively. The last 5-7 orbits are simulated for the different models, all starting from the same orbital separation. 

We performed fully general-relativistic simulations adopting two QHC models, QHC19B (named here QHC19-{\it soft}), QHC19D (named QHC19-{\it stiff}) \cite{baym2019}, and the purely hadronic Togashi EOS \cite{togashi2017}. 
Additional description of the EOSs, the codes, the NS properties, and some of the numerical parameters used in our simulations is presented in the Supplemental Material.
Here, we briefly comment only on how we mimic thermal effects in matter, even when adopting an EOS, like QHC19, that does not contemplate them. Ours is a standard treatment in numerical relativity, but we discuss it nevertheless because it may be of interest to a wider audience. 
Approximate thermal effects are included by adding to the pressure given by the {\it cold} EOS a component calculated by assuming an {\it ideal-gas} behavior with a constant {\it  ideal-gas index} $\Gamma_{\rm th}$, chosen in the range $1.5\text{--}2.0$ to reproduce realistic values (see, {\it e.g.}, \cite{bauswein2010, rezzolla2013, Togashi2014, Lu2019, Figura2020, Huth2021, Raithel2021, Raduta2021, Keller2021}).
Note that the lifetime before collapse to black hole of the material
object formed in the merger depends also on thermal support and thus on
the {\it ad hoc} value of $\Gamma_{\rm th}$, but post-merger oscillation
frequencies (see below) are relatively insensitive to it
\cite{takami2015}. The lifetime before collapse is a quantity that anyway
cannot currently be estimated accurately in numerical simulations,
because it depends sensitively on many factors, including non-physical
ones like grid setup and resolution. We focus, instead, on post-merger
oscillation frequencies and, in order to have higher power in the
oscillation modes, we chose the highest reasonable value, $\Gamma_{\rm
  th} = 2$, which gives the longest lifetime before collapse. See
Sec. \uppercase\expandafter{\romannumeral4} of the Supplemental Material for details.

{\it Results and discussion.}
As seen in Fig.\ref{fig:cs2-nb}, both QHC19-{\it soft} and QHC19-{\it
  stiff} are stiffer (have higher sound speed) than the Togashi EOS at
densities slightly above $2n_0$. The Togashi EOS is stiffer than
QHC19-{\it soft} for $n \gtrsim 3.5n_0$, and stiffer than QHC19-{\it stiff} for $n \gtrsim 4.0n_0$. Within the density range reached in our BNS simulations ({\it cf.} Fig. \ref{fig:n_max}), QHC19-{\it stiff} is thus always stiffer than the Togashi EOS for all models with different masses considered here, while QHC19-{\it soft} is softer, in some regions, for high-mass BNSs. 

In QHC19-{\it stiff}, the sound speed (and thus pressure support) around $3.5n_{0}$ increases the most; it is then expected that inspiraling stars and merged objects in BNSs with QHC19-{\it stiff} are less compact than those with the Togashi or QHC19-{\it soft} EOSs, as can be ascertained in Fig. \ref{fig:n_max}: the maximum number density $n_{\max}(t)$ is smaller than for the other EOSs, in the inspiral, after the merger, and (on average) during the merger. Even in our most massive case, $n_{\max}$ for QHC19-{\it stiff} reaches up only to $\approx 3.8n_{0}$. At such densities, indeed, stiffening due to the crossover is still important. 

In QHC19-{\it soft}, in contrast, the evolution of $n_{\max}$ is
different for binaries of different masses. Since for densities $\lesssim
3.5n_0$ QHC19-{\it soft} is stiffer than the Togashi EOS, in our
lowest-mass case, M1.25, in which densities higher than $3.5n_0$ are
reached only toward the end of our simulations, we see that $n_{\max}$ is always smaller than that for the Togashi EOS. For M1.30, where the maximum density after the merger reaches $3.5\text{--}4n_0$, the differences between the QHC19-{\it soft} and Togashi EOS  appear to average out (their sound-speed curves cross around $3.5 n_0$; {\it cf.} Fig. \ref{fig:cs2-nb}), leading to similar evolution. For even larger masses, M1.35 and M1.375, during and after the merger, densities greater than $\sim 3.5n_0$ are reached in a wide region, and hence QHC19-{\it soft} leads to a considerably more compact merged object.

\begin{figure*}
\centering
\includegraphics[width=1.1\textwidth]{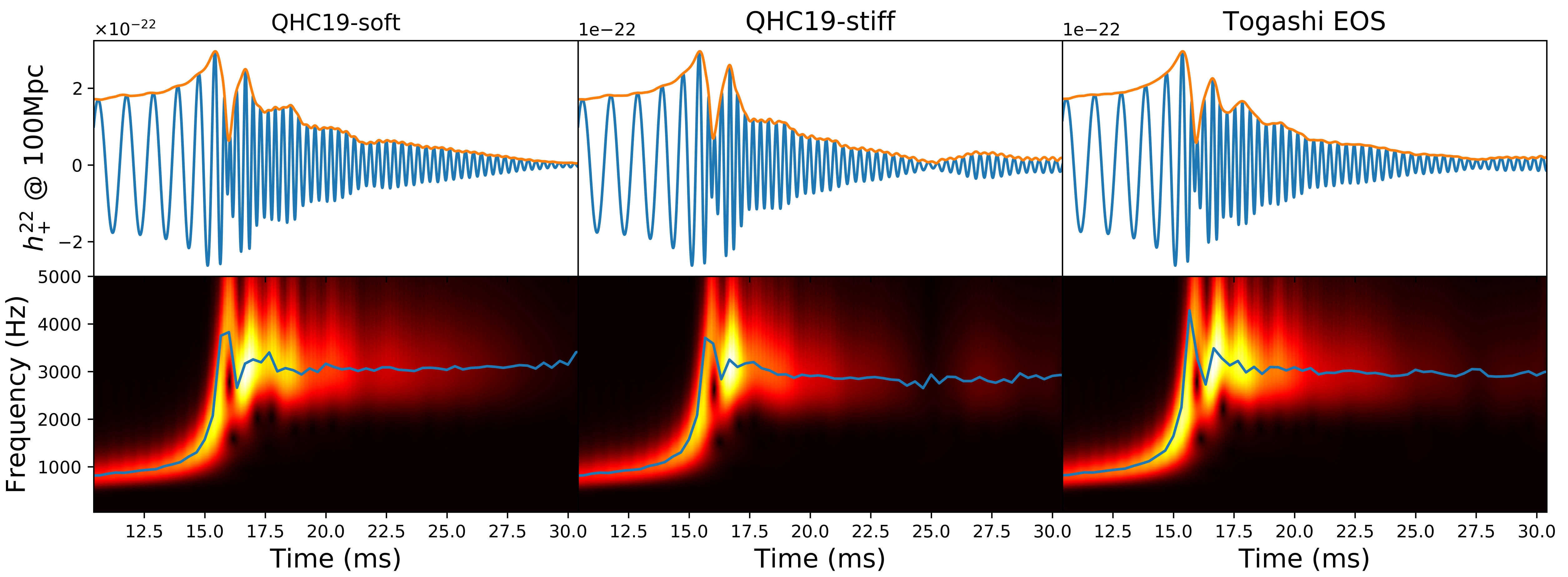}
\caption{Top: Fundamental and dominant harmonic mode ($\ell=m=2$) of the plus polarization of the GW strain with amplitude envelope for the M1.35 models. Bottom: Spectrogram
(brighter colors indicate higher power in the spectrum) 
and instantaneous frequency of the same models.
}
\label{fig:time-frequency}
\end{figure*}

The oscillations of the merged object produce intense GW emission,
characterized by distinct peaks in the power spectrum, whose frequencies
are found to correlate with stellar properties like compactness, average
density, or tidal deformability \cite{Luca2008, bauswein2012a,
  hotokezaka2013, takami2014, takami2015, bernuzzi2015}. At least three
peaks ($f_{1}$, $f_{2}$, and $f_{3}$, sometimes referred to with different names in the literature) may be identified in most cases, but basically only one, $f_{2}$, is not transient and remains even after a few milliseconds \cite{rezzolla2016, Bauswein2016EPJA, Bauswein2019JPhysG, baiotti2019, Friedman2020, Bernuzzi2020, Dietrich2021}. The $f_{2}$ frequency slightly changes in time, as the density profile changes because of GW emission and angular-momentum transfer from inner parts to outer parts \cite{hanauske2017}. 

Figure \ref{fig:time-frequency} displays the fundamental and dominant harmonic mode ($\ell=m=2$) of the plus polarization of the GW strain, $h_{+}^{22}$, for the M1.35 configurations (top panels) and the corresponding time-frequency evolution and instantaneous frequency (bottom panels). Some similarities and differences between our purely hadronic
and QHC models are apparent. The damping times for post-merger GWs
(signaled by the extinguishing of the red color over the whole frequency
band in the spectrogram) are seen to be dependent on the EOS, and the time interval in which a wide range of frequencies has a lot of power (the time interval in which the spectrogram has a bright band) is shorter for QHC19-{\it stiff}. This means that the transient period between the merger and the time when gravitational radiation settles to a well-identified main frequency, $f_2$, is shorter for QHC19-{\it stiff}.

\begin{figure}[htbp]
\centering
\includegraphics[width=0.48\textwidth]{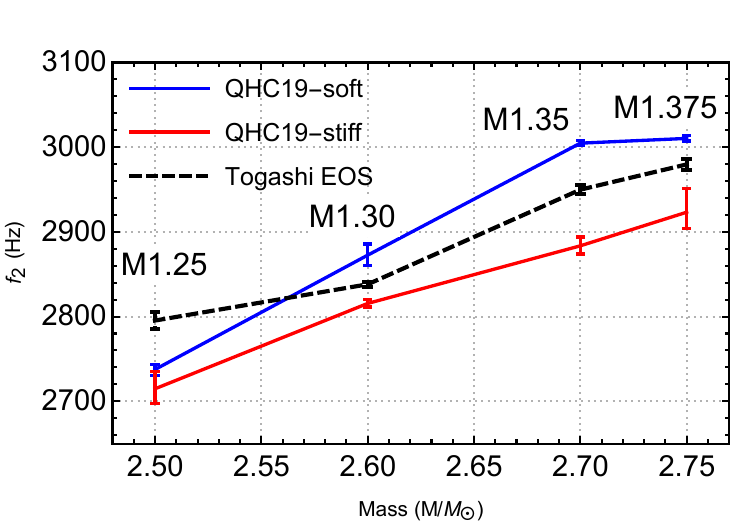}
\caption{Relation between $f_{2}$ and the total mass of the binary. }
\label{fig:f2}
\end{figure}

We also note that in all our simulations (with and without a QHC), the instantaneous frequency in the late post-merger phase after the transient period approximately approaches a constant, though a different one for different models. This is in contrast with hybrid EOS models with 1PTs that predict an abrupt decrease in pressure support, causing the object to shrink rapidly and thus an increase of the instantaneous GW frequency \cite{Bauswein2019, Most:2018eaw, Most2020, weih2020, Blacker2020, Liebling2021, Prakash2021}.  

To estimate quantitatively the peak frequencies together with their
uncertainty range, we employ a Markov Chain Monte Carlo fitting method
\cite{Foreman_Mackey_2013} based on Bayesian inference
\cite{gelmanbda04}. We fit the $f_{1}$ peak with a Gaussian model and the
$f_{2}$ peak with a model that considers skewness. The latter has been
chosen to describe the decay of the mode and the frequency shift during
the transient phase \cite{takami2015, rezzolla2016} (see also
Sec. \uppercase\expandafter{\romannumeral3} of the Supplemental Material).

As is well known, the relations between the values of $f_2$ in different
configurations are similar to the respective relations between the values
of $n_{\rm max}$ (on average) since post-merger frequencies are related
to compactness or average density \cite{Luca2008, bauswein2012a,
  hotokezaka2013, takami2014, takami2015, bernuzzi2015}, but comparing
the values of $f_2$ for different EOSs and BNS masses gives important
clues on how to discriminate observationally different types of quark
dynamics in the high-density end of EOSs. Figure~\ref{fig:f2} shows the
fitting results of $f_{2}$ for different EOSs and binary masses, with
their 68\% fitting uncertainty, which is comparable to the numerical
accuracy of our simulations (see
Sec. \uppercase\expandafter{\romannumeral2} of the Supplemental Material). 


For all masses, $f_2$ for QHC19-{\it stiff} is lower than that for the Togashi
EOS, and this is related to the lower compactness of the merged object,
which is, in turn, related to the pronounced peak in sound speed for
QHC19-{\it stiff}. For QHC19-{\it soft}, except for our lowest-mass case,
$f_{2}$ is higher than that for the Togashi EOS. In models M1.25, $f_{2}$
for both QHC19-{\it soft} and QHC19-{\it stiff} is lower than that for
the Togashi EOS. This is because quark-matter densities ($\sim 5n_0$), where the QHC EOSs are softer than the Togashi EOS, are not reached, and thus the remnant is less compact. This is a unique feature of the peak in sound speed present in QHC models and is independent of the height of such peak (namely, of the parameters of the specific QHC EOS). Note, however, that, since the stiffening in the crossover domain is strongly affected by the quark-matter EOS it is attached to, even in lower-mass models one may still, in principle, gain from observations useful information on how quarks are liberated in high-density hadronic matter.

In order to study further whether it may be possible to discriminate observationally between EOSs with a QHC or with a 1PT, we define $\Delta f_2$ as the difference between the $f_2$ resulting from an EOS with a 1PT or crossover and the $f_2$ resulting from its baseline EOS: $\Delta f_2\equiv f_2^{\rm{\ phase\ transition\ or\ crossover}} - f_2^{\rm{\ baseline}}$. 

For the QHC EOSs employed here, $\Delta f_2$ is in the range $\pm (50-100) \, \rm Hz$
and, more importantly, is negative for all QHC19-{\it stiff} models and
for the lower-mass model of QHC19-{\it soft}. This is in contrast to the
case of EOSs with a 1PT, in which $\Delta f_2$ is always found to be
positive in the literature (see also
Sec. \uppercase\expandafter{\romannumeral5} of the Supplemental Material). This is a qualitative feature that makes relatively simple to discriminate observationally between these different types of EOSs.
In particular, an observation of a low-mass BNS system, as our M1.25
model, would allow one to distinguish between QHC EOSs and EOSs with a 1PT, according to the sign of the measured $\Delta f_2$.

For higher masses and for (weak) 1PTs that result in a $\Delta f_2$ comparable to that of QHC EOSs, it may be difficult to discriminate from observations, unless the 1PT occurs some time after the merger. In this case, the value of $f_{2}$ would change abruptly \cite{weih2020} and, if this change can be measured, it would be a clear difference with respect to QHC EOSs ({\it cf.} Fig. \ref{fig:time-frequency}).

{\it Conclusions.} In this Letter, we performed the first (and fully general-relativistic) simulations of BNS mergers with EOSs based on QHC (QHC19) and discussed how they could be distinguished from purely hadronic EOSs or hybrid quark-hadron EOSs with 1PTs. 

We found that a QHC EOS with a pronounced peak in sound speed, like QHC19-{\it stiff}, leaves a clear and unique signature in the post-merger main frequency: for any binary mass, $f_2$ is lower than that of the baseline hadronic EOS, and thus also lower than that expected for EOSs with a 1PT. 
In higher-mass mergers with the QHC19-{\it soft} EOS, instead, it may be difficult to discriminate from a weak 1PT, unless the value of $f_2$ is observed to change rapidly in time, a signature  of a 1PT occurring after the merger \cite{weih2020}.

Results of this Letter will become relevant to observations when GWs in
the kilohertz band are surveyed with higher sensitivity by upgraded
detectors \cite{LVKprospects2020} and third-generation observatories ({\it e.g.}, the Einstein Telescope \cite{punturo10} and Cosmic Explorer \cite{abbott17}), also with a specifically optimized design ({\it e.g.}, NEMO \cite{ackley20}).

In fact, sensitivities on the order of $50 \, \rm Hz$ in this band, sufficient to distinguish a QHC EOS from a purely hadronic one, are estimated to be reached in these detectors.
For example, Ref. \cite{Chatziioannou2017} estimated that $f_2$ can be
measured to within about 36 (27)\{45\} Hz at the 90\% credible level for
a stiff (moderate) \{soft\} EOS at a post-merger signal-to-noise ratio of
5. Other works make similar predictions, for signal-to-noise ratio
$\gtrsim 10$ \cite{Breschi2019}. A signal-to-noise ratio $\gtrsim 5 - 10$ is predicted to be attainable easily for sources at 200 Mpc or even more by Cosmic Explorer \cite{Srivastava2022} and Einstein Telescope \cite{punturo10}, leading to reasonably frequent measurements.

This work is a first attempt to study in BNS mergers the unique features
of QHC EOSs. We plan to extend the analysis in several directions, first
of all by adopting the QHC21 \cite{kojo2021b} EOS, which improves further
over QHC19 under the microscopical point of view and which was made
public after we finished our simulations. We will explore the
relationship between some EOS parameters and observable quantities, as
well as finite-temperature effects, expected to be important for the
onset of quark saturation \cite{kojo2021a}. 
We also plan to perform simulations of unequal-mass binaries and study the influence of QHC EOSs on mass ejecta.

\acknowledgments

This work is supported in part by the Japan Society for the Promotion of Science (JSPS;  KAKENHI Grants No. 
JP19H00693,  
 No. JP19KK0354,  
 No. JP20H04753,  
 No. JP21H01088,  
 No. T18K03622, 
 No. JP17K14305, 
 No. JH18H03712, 
 No. JH18H05236), 
by the Pioneering Program of RIKEN for Evolution of Matter in the Universe (r-EMU),
by the Graduate Program on Physics for the Universe at Tohoku University, 
and by the National Natural Science Foundation of China (No. 11933010,
No. 11875144, and No.12233011). 
Simulations were performed on the Hokusai supercomputer at RIKEN, on the Aterui II supercomputer (Cray XC50) at the Center for Computational Astrophysics of NAOJ, and on the BSCC-M supercomputer at the Beijing Super Cloud Computing Center.

\bibliographystyle{apsrev4-2}
\bibliography{ref}
\clearpage

\end{document}


\title{Merger and post-merger of binary neutron stars with a quark-hadron crossover equation of state (Supplemental Material)}

\author{Yong-Jia Huang$^{1,2,3}$, Luca Baiotti$^{4}$, Toru Kojo$^{5,6}$, Kentaro Takami$^{7,3}$, Hajime Sotani$^{8,3}$, Hajime Togashi$^6$, Tetsuo Hatsuda$^3$, Shigehiro Nagataki$^{3,8}$ and Yi-Zhong Fan$^{1,2}$}
\noaffiliation

\affiliation{
Key Laboratory of Dark Matter and Space Astronomy, Purple Mountain Observatory, Chinese Academy of Science, Nanjing, 210023, China.}
\affiliation{
School of Astronomy and Space Sciences, University of Science and Technology of China, Hefei, Anhui 230026, China.}
\affiliation{RIKEN Interdisciplinary Theoretical and Mathematical Sciences Program (iTHEMS), RIKEN, Wako 351-0198, Japan.}
\affiliation{International College and Graduate School of Science, Osaka
  University, 1-1 Machikaneyama-cho, Toyonaka 560-0043, Japan}
\affiliation{Key Laboratory of Quark and Lepton Physics (MOE) and Institute of Particle Physics, Central China Normal University, Wuhan 430079, China}
\affiliation{Department of Physics, Tohoku University, Sendai 980-8578, Japan}
\affiliation{Kobe City College of Technology, 651-2194 Kobe, Japan}
\affiliation{RIKEN Astrophysical Big Bang Laboratory (ABBL), Cluster for Pioneering Research, Wako, Saitama 351-0198, Japan}
\maketitle

\section{Numerical setup} 

We have generated initial data for quasiequilibrium irrotational BNSs at a separation of $45\,\mathrm{km}$ (which leads to about 5 - 7 orbits before merger) using the open-source code {\tt Lorene} \cite{Gourgoulhon2001}. We have performed fully general-relativistic hydrodynamic simulations using the {\tt WhiskyTHC} code \cite{radice2014a,radice2014b}, which is written in the {\tt Einstein Toolkit} framework \cite{einsteintoolkit}. In particular, we have employed a finite-volume scheme with $5^{\rm th}$-order monotonicity-preserving reconstruction \cite{Suresh1997} and the Harten-Lax-van Leer-Einfeldt (HLLE) Riemann solver \cite{hlle1983}. The spacetime evolution is calculated in the Z4c formulation \cite{bernuzzi2010} through the {\tt CTGamma} code, with ``$1+\log$'' slicing and ``Gamma-driver'' shift conditions \cite{alcubierre2003,pollney2007}. For the time integration of the coupled set of the hydrodynamic and Einstein equations we have used the method of lines, with a third-order strong-stability-preserving Runge-Kutta scheme \cite{gottlieb2009} with a Courant-Friedrichs-Lewy~(CFL) factor of $0.075$ (such a small value is necessary when adopting flux reconstruction in local-characteristic variables using the adopted monotonicity-preserving scheme \cite{Suresh1997}). The simulation grids with adaptive-mesh refinement are managed through the {\tt Carpet} code \cite{schnetter2004}. The simulation domain extends to $\approx 1477\,{\rm km}$, and we use seven mesh-refinement levels with the finest grid spacing $\approx 231\,{\rm m}$ for our fiducial simulations.

\section{Numerical accuracy}

\begin{figure}[htbp]
    \centering
    \includegraphics[width=0.45\textwidth]{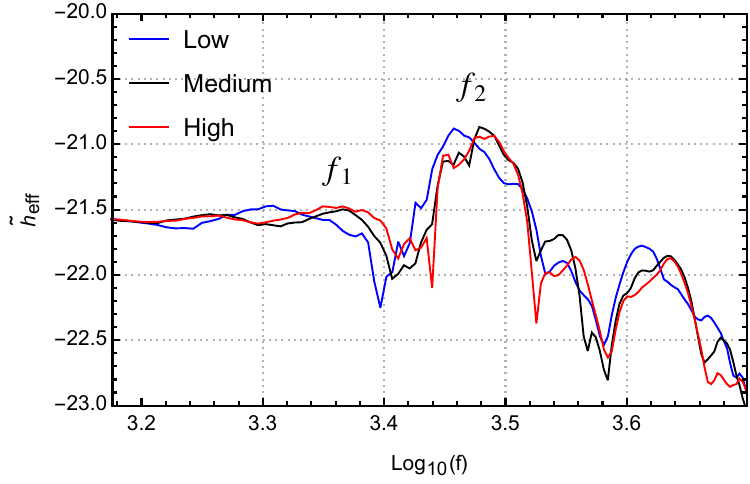}
    \caption{The GW spectrum of the Togashi EOS for the M1.25 configuration with low ($\Delta x\approx 369$ m, blue), medium ($\Delta x\approx 231$ m, black), and high ($\Delta x\approx 185$ m, red) resolution, respectively. }
    \label{fig:resolution}
\end{figure}

We have checked the dependence of $f_2$ on numerical resolution. Figure~\ref{fig:resolution} shows the power spectral density (PSD) of the GW signal for the M1.250 models with the Togashi EOS with three resolutions: Low (finest grid spacing $\Delta x\approx 369$ m), Medium ($\Delta x\approx 231$ m) and High ($\Delta x\approx 185$ m). While the difference in $f_2$ between the Low and Medium resolutions is somewhat large, the one between the Medium and High resolutions is smaller ($\sim 20 \, \rm Hz$) and comparable with the error resulting from the fitting procedure. Therefore, we adopt the Medium resolution as our fiducial one.

\section{Fitting of the GW spectrum}\label{sec:fitting}

We give here details about our analysis of the GW spectrum. We considered the procedure employed in \cite{takami2015} and tried to improve it further in several directions. Here is a list of the main differences. The procedure itself will be described in detail further below.

\begin{table*}[htbp]
\caption{Properties of the BNS merger systems simulated in this work. The quantities in the table are, from left to right, gravitational mass $M$ of one of the NSs when in isolation, baryon mass $M_{\rm b}$ of one NS, radius $R$ of one of the NSs when in isolation, Arnowitt-Deser-Misner (ADM) mass $M_{\rm ADM}$ of the BNS system at the initial time, total angular momentum $J$ at the initial time, orbital frequency $f_{\rm orb}$ at the initial time, dimensionless tidal deformability $\Lambda = \frac{2}{3}k_{2}(R/M)^5$ where $k_{2}$ is the tidal love number \cite{hinderer2007,hinderer2009}, contact frequency $f_{\rm cont} = C^{3/2}/(2\pi M)$, where $C$ is compactness, post-merger main frequency $f_{2}$, described by model parameter $F_{2}$ in Eq. (S6) with its 68\% confidence interval, and the mean frequency of the $f_{2}$ peak $f_{\rm{2, mean}}$.}
\begin{ruledtabular}
\begin{tabular}{lcccccccccl}
Model & $M$[$M_{\odot}$] & $M_{\rm b}$[$M_{\odot}$] & $R[{\rm km}]$ & $M_{\rm ADM}$[$M_{\odot}$] & $J$[$M_{\odot}^{2}$] & $f_{\rm orb}$[Hz] &$\Lambda$ &$f_{\rm cont}$[Hz]  & $f_{2}$[Hz] & $f_{\rm 2, mean}$[Hz]\\
\hline
QHC19-{\it soft}-M1.25 & 1.250 & 1.37 & 11.58 & 2.47 & 6.39 & 272.92 & 608.24& 1646.87 & $2738^{+6}_{-7}$ & 3119\\
QHC19-{\it soft}-M1.30 & 1.300 & 1.43 & 11.58 & 2.57 & 6.82 & 277.26 & 483.85& 1744.62 & $2873^{+13}_{-12}$ & 3187\\
QHC19-{\it soft}-M1.35 & 1.350 & 1.49 & 11.59 & 2.67 & 7.26 & 281.50 & 386.72& 1844.80 & $3005^{+3}_{-3}$ & 3271\\
QHC19-{\it soft}-M1.375 & 1.375 & 1.52 & 11.59 & 2.72 & 7.44 & 283.45 & 346.02& 1895.79 & $3010^{+3}_{-3}$ & 3273\\
QHC19-{\it stiff}-M1.25 & 1.250 & 1.37 & 11.58 & 2.47 & 6.39 & 272.97 & 608.79& 1646.65 &$2715^{+20}_{-17}$ & 3046\\
QHC19-{\it stiff}-M1.30 & 1.300 & 1.43 & 11.59 & 2.57 & 6.82 & 277.27 & 484.72& 1744.17 & $2816^{+5}_{-4}$ & 3177\\
QHC19-{\it stiff}-M1.35 & 1.350 & 1.49 & 11.60 & 2.67 & 7.26 & 281.39 & 387.99& 1843.85 & $2884^{+10}_{-10}$ & 3195\\
QHC19-{\it stiff}-M1.375 & 1.375 & 1.52 & 11.60 & 2.72 & 7.49 & 283.42 & 347.61& 1894.57 & $2923^{+28}_{-19}$ & 3213\\
Togashi-EOS-M1.25 & 1.250 & 1.37 & 11.57 & 2.47 & 6.39 & 273.00 & 603.32& 1647.29 & $2795^{+11}_{-10}$ & 3136\\
Togashi-EOS-M1.30 & 1.300 & 1.43 & 11.58 & 2.57 & 6.82 & 277.27 & 478.87& 1745.75 & $2838^{+3}_{-3}$ & 3190\\
Togashi-EOS-M1.35 & 1.350 & 1.49 & 11.58 & 2.67 & 7.26 & 281.39 & 381.75& 1846.71 & $2950^{+5}_{-5}$ & 3257\\
Togashi-EOS-M1.375 & 1.375 & 1.52 & 11.58 & 2.72 & 7.49 & 283.39 & 341.27& 1898.25 &  $2980^{+7}_{-7}$ & 3254\\

\end{tabular}
\end{ruledtabular} 
\label{tb1}
\end{table*}

First, we have employed Bayesian inference \cite{gelmanbda04} and the Markov-chain-Monte-Carlo (MCMC) method \cite{Foreman_Mackey_2013} in the estimation of parameters describing the GW spectrum. With this method, we can estimate the optimal value and its confidence interval for each parameter. MCMC also allows for a more efficient search in a large parameter space, avoiding local extrema in the optimization of fitting results.

Second, we choose an alternative treatment for removing the contribution of the inspiral to the post-merger GW spectrum. To limit the uncertainty resulting from including in the MCMC fitting the very-low-frequency unphysical peak arising from the high-pass filter used in \cite{takami2015} ({\it cf.} Fig. 3 of \cite{takami2015}), we have tried to single out the GW contribution of the inspiral and removed it before fitting the post-merger peaks.
By inspecting power spectra of BNS systems that do not show clear post-merger peaks (like those in Fig.6 of \cite{read2013}), we have noted that the contribution of the inspiral to the power spectrum could be approximately described by a power-law at lower frequencies and an exponential decay at higher frequencies (the post-merger part). We then use such a power-low-plus-exponential-decay profile to filter out the inspiral contribution before fitting the post-merger peaks.

Third, although the three post-merger peaks in the GW spectrum, $f_{1}$, $f_{2}$ and $f_{3}$, can be described, for example, with a toy model based on a harmonic oscillator embedded in a rigidly rotating disk \cite{takami2015}, there are currently no analytical clues about the shape and width of the peaks.
Therefore, we adopt a simple model with as few parameters as possible to avoid overfitting.
In practice, in our analysis we focus on the $f_{1}$ and $f_{2}$ peaks and use a Gaussian function for fitting $f_{1}$ and a skewed Gaussian function for fitting $f_{2}$, since the $f_{2}$ peak is usually asymmetric as the frequency of this oscillation changes slightly after the merger (especially in the first few milliseconds). A skewed Gaussian function has an additional parameter that describes the shape of the curve. In total, seven parameters are used.

Hereafter, we summarize the whole fitting procedure. 

\begin{enumerate}
    \item We use a Tukey window function \cite{Harris78, McKechan2010, takami2015} with a factor of 0.25 to reduce the non-physical noise from waveform truncation.
    
    \item We calculate PSD of the root-mean-square amplitude $\tilde{h}$ for the whole duration of our simulations, namely, from the initial orbital separation of $45$ km to $19$ ms after the merger time, defined as the time of the maximum amplitude of the GW amplitude.
    
    \begin{equation}
        \tilde{h}(f) \equiv \sqrt{\frac{\left| \tilde{h}_{+}(f) \right|^2 + \left| \tilde{h}_{\times}(f) \right|^2}{2}}
    \end{equation}
    
    where the $\tilde{h}_{+}$ and $\tilde{h}_{\times}$ are the Fourier transforms of the two polarization modes ($h_{+}$ and $h_{\times}$) of the GW with a 30 Hz high-pass filter, namely,
    
    \begin{equation}
        \tilde{h}_{+, \times}(f) \equiv\left\{\begin{array}{ll}
\int h_{+, \times}(t) e^{-i 2 \pi f t} d t & (f \geq 30 \, \rm{Hz}) \\
0 & (f < 30 \, \rm{Hz})
\end{array}\right.
    \end{equation}
    
    In this work, we consider only the $l = m =2$ mode, which is the dominant one both in the inspiral and after the merger:

    \begin{equation}
        h_{+, \times}=\sum_{\ell=2}^{\infty} \sum_{m=-\ell}^{\ell} h_{+, \times}^{\ell m}  Y_{\ell m}^{(s)}(\theta, \varphi) \approx h_{+, \times}^{22}Y_{22}^{(s)}(\theta, \varphi),
    \end{equation}

    which represents the expansion of the GW strain in terms of the $s$ spin-weighted spherical harmonics, $Y_{\ell m}^{(s)}(\theta, \varphi)$.
    Here $s=-2$.
     \cite{Goldberg1967}.

    \item We fit the contribution of the inspiral to the post-merger GW spectrum using a power-law function at low frequencies, with exponential damping starting at the contact frequency $f_{\rm cont} = C^{3/2}/(2\pi M)$ (where $C$ is compactness). 
    
    \item
    We perform Bayesian inference to determine the peak parameters with their uncertainty (68\% confidence interval). The posterior probability of the parameters $\theta_{i}$ of the fitting model can be written as:

\begin{equation}
P(\theta_{i}|2\tilde{h}(f)f^{1/2}, H) = \frac{P(2\tilde{h}(f)f^{1/2}| \theta_{i},H)P(\theta_{i}|H)}{P(2\tilde{h}(f)f^{1/2}|H)},
\label{eq:posterior}
\end{equation}

where $2\tilde{h}(f)f^{1/2}$ is the PSD of the strain, and $H$ represents the fitting model. The parameters in the prior  $P(2\tilde{h}(f)f^{1/2}|H)$ are uniformly sampled in the allowed parameter space. The likelihood function can be written as:


\begin{eqnarray}
\begin{aligned}
&  \ln P(2\tilde{h}(f)f^{1/2}|\theta_{i}, H) = & \\ 
&  -\frac{1}{2}\sum_n \left[\frac{(S(\theta_{1},...,\theta_{i},f_n) - 2\tilde{h}(f_n)f_n^{1/2})^2}{s_n^2}+ \ln (2\pi s_n^2)\right] . &  
\end{aligned}
\label{eq:likelihood}
\end{eqnarray}

It gives the expected probability with the given set of parameters in the model $S(\theta_{1},...,\theta_{i},f_n)$ (for simplicity, hereafter, we use the symbol $S(f)$ instead of $S(\theta_{1},...,\theta_{i},f_n)$). 
In practice, we perform the MCMC fitting via the $\it{emcee}$ code \cite{Foreman_Mackey_2013}, and the variance $s_n^2$ is estimated by $\it{emcee}$ during the fitting.
\end{enumerate}

For step 3, we describe the contribution of the inspiral to the GW spectrum $\tilde{h}_{\rm insp}(f)$ as follows:
\begin{equation}
\label{eq:h_insp}
\tilde{h}_{\rm insp}(f) =\left\{
\begin{array}{lcl}
c_{0}{\rm Log}_{10}f + c_{1},      &     & { f < f_{\rm cont} }  \\
 c_{0}{\rm Log}_{10}f + c_{2} + c_{3}f,    &     & { f \geq f_{\rm cont}}\\
\end{array} \right. 
\end{equation}
where $c_{i}$ ($i$ = 0, 1, 2, 3) are model parameters. 
Parameter $c_{2}$ is determined by imposing continuity of $\tilde{h}_{\rm insp}(f)$:
\begin{equation}
c_{2} = c_{1} - c_{3}f_{\rm cont}.
\end{equation}

\begin{figure}[htbp]
    \centering
    \includegraphics[width=0.45\textwidth]{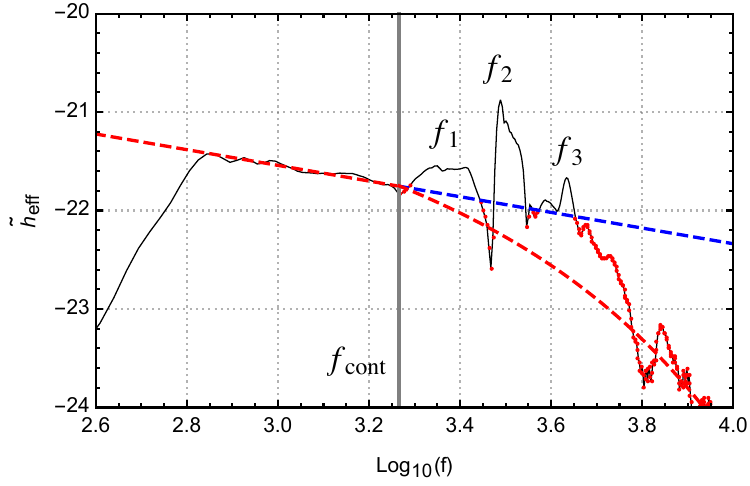}
    \caption{Fitting for the contribution of the inspiral to the 
    GW spectrum, for model QHC19-{\it soft}-M1.35. The black line is the PSD of the whole signal. The red dashed line describes the power-law evolution for the inspiral spectrum and is extended to higher frequencies as a reference (blue dashed line). Exponential decay is assumed to start from the contact frequency (gray vertical line), and the data points (red dots) selected by removing the post-merger peaks are used in fitting the high-frequency contribution of the inspiral.}
    \label{fig:fitting}
\end{figure}

As an example, Fig.~\ref{fig:fitting} illustrates our fitting procedure for the inspiral contribution in the case of the QHC19-{\it soft}-M1.35 configuration. The black line is the PSD of the whole signal and the red dashed line shows the contribution of the inspiral part to the PSD. We assume that a power-law in the PSD of the waveform, $\tilde{h}_{\rm eff}(f)\equiv {\rm Log}_{10}(2\tilde{h}(f)f^{1/2})$, describes the binary inspiral until the contact frequency $f_{\rm cont}$ (gray vertical line). Such power law also provides an upper limit for the inspiral contribution at higher frequencies (blue dashed line). This means that we consider the data points that have lower power than the extension of the power-law describing the inspiral (blue dashed line) as not-belonging to the post-merger. We mark these data points as red dots in the figure, fit them with the power-law + exponential-decay described in Eq. (\ref{eq:h_insp}) (red dashed line), up to $10^{4}\, \rm Hz$, and subtract the result from $\tilde{h}_{\rm eff}(f)$ to obtain a better estimate of the PSD in the post-merger: $\tilde{h}_{\rm post}(f) = \tilde{h}_{\rm eff}(f) - \tilde{h}_{\rm insp}(f)$.

\begin{figure}[htbp]
    \centering
    \includegraphics[width=0.45\textwidth]{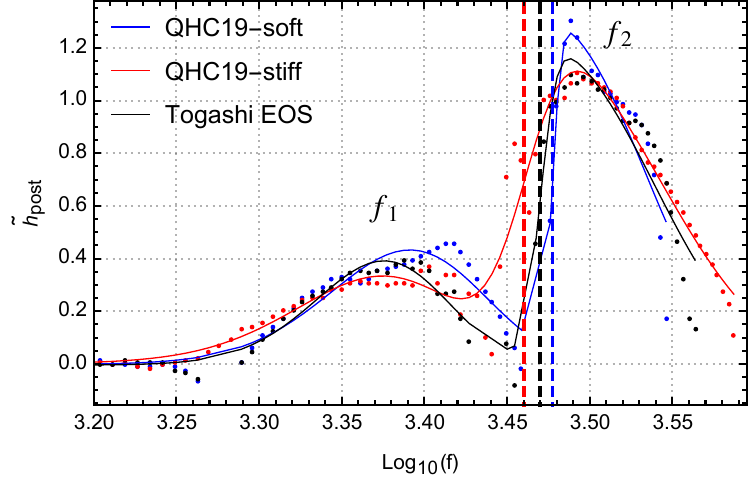}
    \caption{The GW spectrum after removing the higher-frequency contribution of the inspiral for QHC19-{\it soft} (blue), QHC19-{\it stiff} (red) and Togashi EOS (black) for the M1.35 configuration. Dotted, continuous and dashed lines represent, respectively, data points, fitting curves and the optimal value of parameter $F_{2}$ [{\it cf.} eq.~(\ref{eq: S2})], which corresponds to the main post-merger frequency. }
    \label{fig:peaks}
\end{figure}

\begin{figure*}[htbp]
    \centering
    \includegraphics[width=1.0\textwidth]{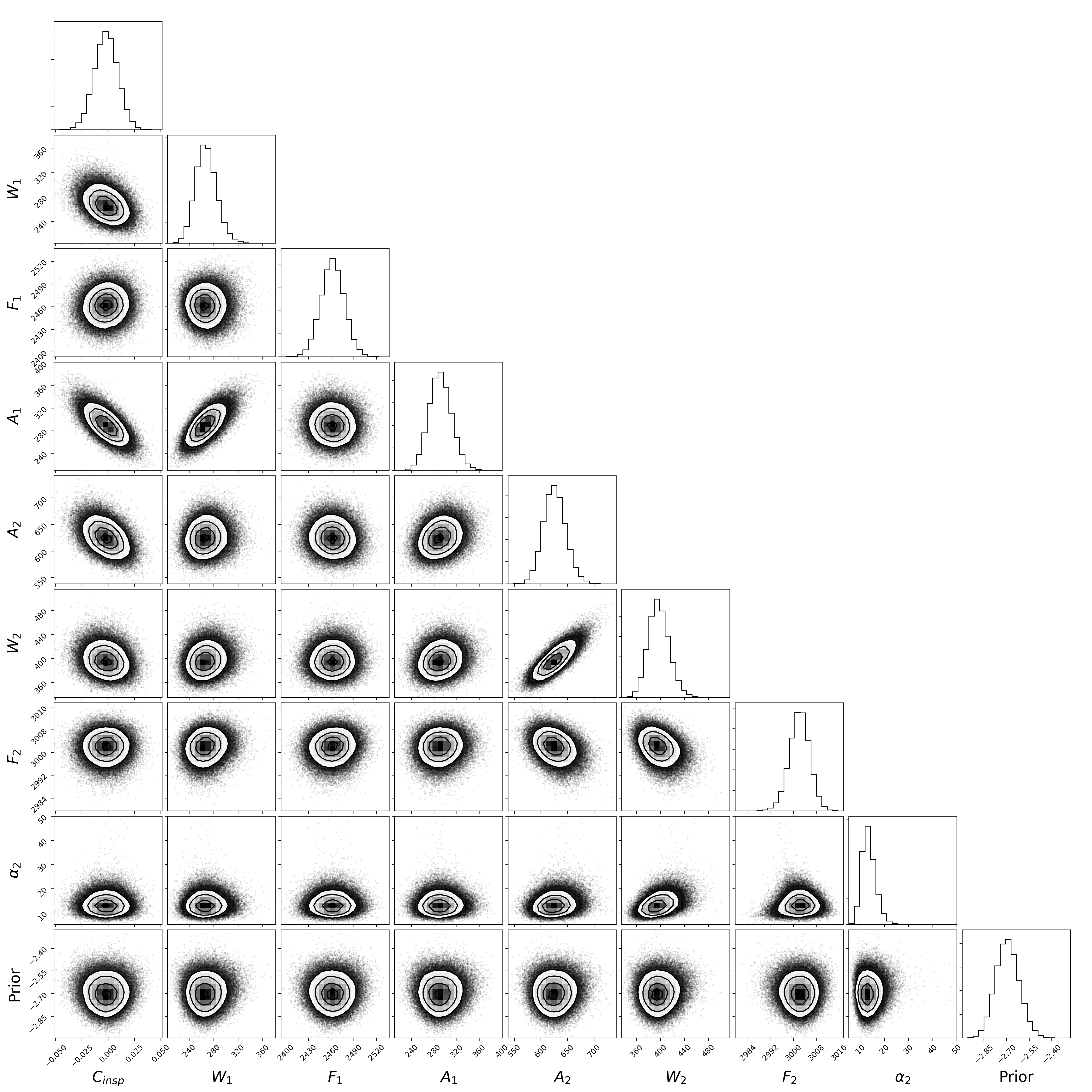}
    \caption{Contour plots for the probability of the fitting results with their probability-density distributions for the QHC19-{\it soft}-M1.35 configuration. Besides the parameters used in the fitting of the PSD, two additional quantities $C_{\rm insp}$ and $Prior$ are shown. They represent the residual of the contribution of the inspiral at high frequencies and the prior distribution of our fitting, respectively. 
    Each parameter converges well to its optimal value.}
    \label{fig:contour}
\end{figure*}

Figure~\ref{fig:peaks} shows the two post-merger peaks of interest ($f_{1}$ and $f_{2}$), after such a removal of the contribution from the inspiral. 
We fit the $f_{1}$ and $f_{2}$ peaks with seven parameters. The fitting function, $S(f)$, that we use for the two peaks is

\begin{equation}
S(f) =  A_{1}e^{-\frac{(f-F_{1})^{2}}{2W_{1}^{2}}} + S_{2},
\label{eq:fitting-main}
\end{equation}
where
\begin{equation}
S_{2} = A_{2}e^{-\frac{(f-F_{2})^{2}}{2W_{2}^{2}}}\int_{-\infty}^{\alpha_{2}\frac{(f-F_{2})}{W_{2}}}e^{-\frac{t^{2}}{2}}dt,
\label{eq: S2}
\end{equation}
and  $A_{i}$, $W_{i}$, $F_{i}$ (${i}=1,~2$) describe the height, width, and location of each peak, respectively. This means that we use a skewed Gaussian model with one more parameter $\alpha_{2}$ to describe the shape of $f_{2}$ \cite{skew-Gaussian}.
In Table~\ref{tb1}, we report the median of the posterior probability of model parameter $F_{2}$ directly, with its uncertainty as the peak frequency $f_{2}$ and show it in Fig.5 of the main text. For a better connection with other works, in Table~\ref{tb1} we also show the mean of $f_{2}$, following the definition of Eq. (21) in \cite{takami2015}, together with other physical quantities of our models. 

We find a first guess for the values (reference values) of the parameters using nonlinear least-squares methods, then extend the parameter space around the reference values and perform MCMC fitting.  Figure~\ref{fig:contour} shows the contours of relevant results for simulation model M1.35 with the QHC19-{\it soft} EOS. 

To test whether a further parameter may still be needed to describe the contribution of the inspiral to the post-merger spectrum, we have also tried to consider a further additive constant, $C_{\rm insp}$, in the right-hand-side of Eq. (\ref{eq:fitting-main}). Fitting results in Fig.~\ref{fig:contour} show that the value of  $C_{\rm insp}$ is less than 1\%, indicating that our scheme for removing the high-frequency contribution of the inspiral is efficient and thus that additional parameters are not required.

\section{Equations of state} \label{sec:EOS}
In this section we give more details on the equations of state (EOSs) that we adopted in our simulations.

\subsection{Quark-Hadron-Crossover equation of state}

Quark-Hadron-Crossover (QHC) EOSs consist in three parts: a nuclear matter EOS at the lowest densities, a quark-matter EOS at the highest densities and, in the region of crossover between hadrons and quark matter, an interpolation that uses the above two EOSs as  boundary conditions \cite{Masuda:2012kf,Masuda:2012ed,Kojo:2021wax,Baym:2019iky,Baym:2017whm}. Here we briefly present the underlying physics of the nuclear and quark-matter EOS.

Nuclear matter in QHC EOSs contains protons, neutrons, electrons, and muons as the effective degrees of freedom. The many-body Hamiltonian used do describe these systems contains empirical nuclear two- and three-body forces which are determined from two-body scattering experiments and the spectroscopy of light nuclei. Electrons and muons are treated as ideal gases but play important roles in maintaining the charge neutrality of the system. For a given baryon density, such a many-body Hamiltonian is used to calculate the energy density of the system, $e(n)$, namely the EOS \cite{Togashi:2017mjp,Akmal:1998cf,Drischler:2020hwi,Drischler:2020yad,Serot:1984ey}. In QHC19 \cite{Baym:2019iky} (adopted in our numerical simulations), we use the Togashi EOS \cite{Togashi:2017mjp} up to $2n_0$. The Togashi EOS is based on variational calculations in which trial many-body wavefunctions are optimized to achieve the minimum energy of the system. The Togashi EOS is similar to the time-honored Akmal-Pandharipande-Ravenhall (APR) EOS \cite{Akmal:1998cf} for the nuclear liquid part but also contains the crust EOS not included in the APR EOS. The Togashi EOS is consistent with nuclear data around $n \simeq n_0$. Moreover, the advantage of using microscopic interactions over phenomenological nuclear-matter models (density functionals arranged to fit the nuclear saturation properties \cite{Serot:1984ey}) is that one can keep tracking the relative importance between two- and three-body potentials, and therefore one can check whether the truncation of forces beyond three-body forces is valid. It turns out that three-body forces are sizable for $\gtrsim 2n_0$ (see, {\it e.g.}, Ref.\cite{Akmal:1998cf}), suggesting that the many-body expansion does not converge. This prevents us from using nuclear EOSs beyond $\sim 2n_0$.

For describing quark matter, we use the Nambu--Jona-Lasinio (NJL) quark
model as a template in which gluonic processes are replaced by effective
quark interactions. The model contains up-, down-, and strange-quarks
\cite{Hatsuda:1994pi} in the QCD part and also includes electrons and
muons for charge neutrality. The model can explain the constituent quark
masses in hadrons (about 1/3 of baryon masses and 1/2 of meson masses) as
dynamically generated masses. The model provides good descriptions for
hadron phenomenology related to the dynamics of the constituent quarks
and to the Nambu-Goldstone bosons ({\it e.g.}, pions) associated with
chiral symmetry breaking. Although the NJL model itself does not have
dynamical effects to confine quarks into hadrons, the model is useful to
describe quark dynamics inside of a hadron whose radius is $\simeq 0.5
\text{--} 0.8$ fm \cite{Manohar:1983md}. In the context of dense matter,
baryons should overlap at $n\simeq 4\text{--}7 n_0$, and then the
constituent quarks begin to directly contribute to the EOS. Here we assume the validity of the NJL descriptions and hence in QHC19
we directly use the NJL estimates on the EOS for $n \gtrsim 5n_0$, with the
addition of two effective interactions which are supposed to become more
important at a higher density. The first is short range density-density repulsion (vector repulsion), whose strength is characterized by the coupling strength $g_{\rm V}$, and the second is attractive diquark correlation, with coupling strength $H$, which drives color superconductivity in the high-density regime. Ref. \cite{Baym:2019iky}, which introduces the QHC19 EOS, explored wide ranges of $g_{\rm V}$ and $H$ and identified the region in the ($g_{\rm V}, H$) plane where there is consistency with constraints obtained from NS observations. Later, Ref. \cite{Song:2019qoh} also studied what are the plausible regions in that plane by matching with microscopic theoretical calculations that manifestly include nonperturbative gluons, and found consistency with NS constraints. While there is still room for quantitative improvements, the importance of nonperturbative physics around $n= 5 \text{--}10 n_0$ is well established, and this is consistent with the breakdown of the weak-coupling calculations for $\lesssim 40 n_0$ \cite{Kurkela:2009gj}.

The EOS in the density range between $\sim 2n_0$ and $\sim 5n_0$ is the most difficult to calculate since the degrees of freedom, baryons or quarks or something intermediate, are not clear-cut. For the EOSs in this density domain, we interpolate the nuclear and quark EOSs in pressure with fifth-order polynomials in the baryon chemical potential $\mu$ \cite{Baym:2017whm}.
The interpolating polynomials are matched with the nuclear and quark-matter EOS boundaries up to second-order derivatives, 
and constrained by imposing thermodynamic stability ($\partial^2 P/\partial \mu^2 \ge 0$) and causality ($\partial P/\partial e = c_{\rm s}^2 \le c^2$ with $c_{\rm s}$ and $c$ being the sound and light speeds, respectively) conditions.

\begin{figure}[htbp]
\centering
\includegraphics[width=0.5\textwidth]{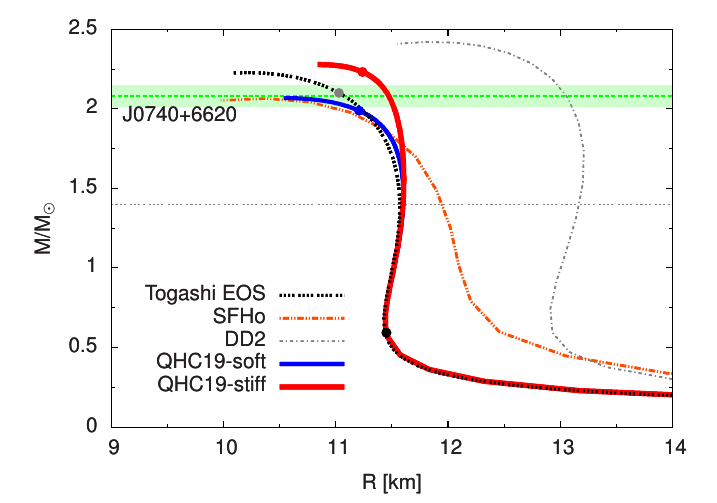}
\caption{
$M$-$R$ relations for pure hadronic EOSs (Togashi EOS, SFHo, DD2), QHC19-{\it soft} and QHC19-{\it stiff}. See also Fig.2 in the main text for the corresponding sound speeds. Two filled circles on each curve display the locations where the maximum density of the NS is $2n_0$ and $5n_0$.The mass of PSR J0740+6620, $M/M_\odot=2.08\pm 0.07$ \cite{Fonseca:2021wxt}, is also shown.
}
\label{fig:M-R}
\end{figure}

The interpolated EOS generally leads to a peak in sound speed, shown in Fig.2 of the main text. The corresponding $M$-$R$ relations are shown in Fig.\ref{fig:M-R} of this Supplemental Material.
In this framework of phenomenological interpolation, the peak in sound speed arises from the need to reconcile soft nuclear EOS at low densities with the $2M_\odot$ constraints  \cite{Masuda:2012kf,Masuda:2012ed,Kojo:2021wax,Baym:2019iky,Baym:2017whm,Bedaque:2014sqa,Tews:2018kmu,Drischler:2020fvz}.
 Recently several works addressed the origin of peaks directly from microscopic considerations \cite{McLerran:2018hbz,Jeong:2019lhv,Kojo:2021ugu,Ma:2019ery}.
The importance of quark substructure for nucleons and of the Pauli blocking effects at quark level is highlighted in all these works.
As discussed in Ref. \cite{Kojo:2021ugu}, quark states at low momenta in a many-nucleon system are saturated at a density not much larger than $n_0$, 
certainly less than the density where baryon cores overlap. 
As density increases the saturated quark levels gradually establish a quark Fermi sea.
At the same time, descriptions of baryons become drastically different from what we would expect from a pure nucleonic picture;
the quark Pauli blocking constraint requires baryons to occupy higher momentum states, pushing baryons from non-relativistic to relativistic regimes \cite{McLerran:2018hbz}.
In this scenario, rapid stiffening and the associated peak in sound speed are triggered by quark saturation effects,
and hence the peak in sound speed can be regarded as the signature of the onset of quark-matter formation \cite{Kojo:2021ugu}.

QHC21 \cite{Kojo:2021wax}, QHC19 \cite{Baym:2017whm}, and QHC18 \cite{Baym:2019iky} are available at the CompOSE archive at \url{https://compose.obspm.fr}.

Note that in this work we adopted only two of the four QHC models proposed in \cite{baym2019}, which correspond to different EOS parameter sets (named A, B, C, D in \cite{baym2019}), relative to the way of connecting the hadron and quark EOSs. Set A is not discussed here because it leads to an EOS resulting in a maximum mass for NSs that is smaller than the mass of some observed NSs. Among the remaining three sets, for simplicity we have chosen only two: the softest (model B) and stiffest (model D) ones in the crossover region.

\subsection{Thermal part of the equation of state}

As mentioned in the main text and as often done in numerical-relativity simulations adopting cold EOSs, we mimic thermal effects by adding to the pressure given by the cold EOS a component calculated approximately (and not taking into account the composition of matter) by assuming an {\it ideal-gas} behavior with a constant {\it ideal-gas index} $\Gamma_{\rm th}$, which is chosen in the range $1.5 \text{--} 2.0$ to reproduce somewhat realistic values (see, {\it e.g.}, \cite{bauswein2010, rezzolla2013, Togashi2014, Lu2019, Figura2020, Huth2021, Raithel2021, Raduta2021, Keller2021}). Also note that the constant $\Gamma_{\rm th}$ as defined here cannot be larger than 2, because that would violate causality (see, {\it e.g.}, \cite{rezzolla2013}).

More in detail, the total pressure $P$ and specific internal energy $\varepsilon$ (note that the relation between energy density and specific internal energy is $e = (\varepsilon + c^2) \rho$, where $\rho$ is the rest-mass density) are set to be the sum of a cold and a thermal component,  $P = P_{\rm c} + P_{\rm th}$ and $\varepsilon = \varepsilon_{\rm c} + \varepsilon_{\rm th}$, where the cold components are from the adopted EOS table and the thermal component for pressure is obtained as $P_{\rm th} = \rho\varepsilon_{\rm th}(\Gamma_{\rm th}-1)$. From $\rho$ and $\varepsilon$ numerically computed from the hydrodynamics equations at a given time step, $P_{\rm c}$, $\varepsilon_{\rm th}$ and $P_{\rm th}$ are calculated. Finally, $P$ is calculated as the sum of $P_{\rm c}$ and $P_{\rm th}$.

\section{Comparison with equations of state with a first-order phase transition}

As mentioned in the main text, several works \cite{Most:2018eaw, Most2020, Bauswein2019, Blacker2020, weih2020, Liebling2021, Prakash2021} have studied the GW signal from BNS mergers described by an EOS containing a first-order phase transition. 
In general, it is thought and observed in simulations that, after a first-order phase transition has taken place after the merger, the $f_{2}$ frequency is higher than that expected from the inspiral signal, which is produced before its occurrence \cite{Bauswein2019, weih2020}.
Here, we want to present a qualitative comparison between the EOSs with a QHC used in our simulations and EOSs with a first-order phase transition. In particular, we highlight how the $f_2$ frequency is different from that expected from the baseline hadronic EOS, using the quantity $\Delta f_2$ already defined in the main text:
\begin{equation}
\Delta f_2\equiv f_2^{\rm{\ phase\ transition\ or\ crossover}} - f_2^{\rm{\ baseline}}   \ .
\end{equation}

We compare our results with those obtained in other works that used the DD2F-SF7  \cite{Bauswein2019, Blacker2020} EOS, which has the DD2F EOS \cite{Typel2005, Typel2010, AlvarezCastillo2016} as hadronic baseline, or the BLQ EOS \cite{Prakash2021}, which has the BLh EOS \cite{Logoteta2021} as hadronic baseline.

\begin{figure}[htbp]
    \centering
    \includegraphics[width=0.45\textwidth]{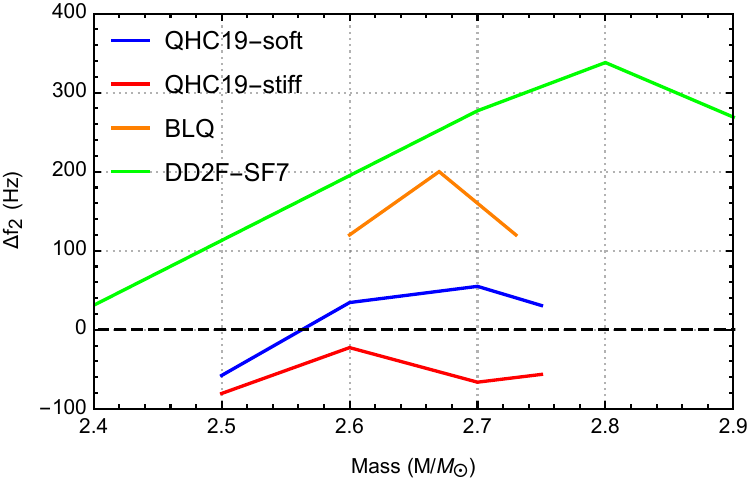}
    \caption{Difference between the $f_{2}$ frequency for models with a QHC (QHC19-{\it soft}, blue, QHC19-{\it stiff}, red) or a first-order phase transition (DD2F-SF7 \cite{Blacker2020}, orange, BLQ \cite{Prakash2021}, green) with respect to the $f_{2}$ frequency for their respective baseline hadronic EOS, for different BNS masses. Data taken from \cite{Blacker2020, Prakash2021}.}
    \label{fig:delta_f2}
\end{figure}

Note that this comparison can only be qualitative for two main reasons. First, the hadronic baseline EOSs in the EOSs with a first-order phase transition are different from the one we employed (Togashi EOS). 
Second, the procedure to extract the $f_2$ frequency from simulation data is different.

Keeping the above caveats in mind, in Fig.~\ref{fig:delta_f2} we show $\Delta f_2$ for the two QHC models that we discuss in the main text, and for DD2F-SF7 and BLQ.
The most notable difference is that, for any BNS mass, $\Delta f_2$ is positive for the EOS with a first-order phase transition, while for both QHC models $\Delta f_2$ is negative for the lowest binary mass that we have simulated (for QHC-19-{\it stiff} it is negative for all binary masses). This is because of the rapid stiffening in QHC EOS, to be contrasted with the softening of the EOS with a first-order phase transition.

Note that here we only focus on the sign of $\Delta f_2$, while the quantitative differences are more model dependent. Choosing different models for EOSs with a first-order phase transition among those of \cite{Bauswein2019, Blacker2020, Prakash2021}, would only change Fig.~\ref{fig:delta_f2} quantitatively ($\Delta f_2$ may be hundred of Hz in some models of first-order phase transition), but not qualitatively.

\clearpage
\bibliographystyle{apsrev4-1}
\bibliography{ref-supplemental}